# Topological insulator $Bi_2Te_3$ films synthesized by metal organic chemical vapor deposition

Helin Cao[1,2], Rama Venkatasubramanian[3,*], Chang Liu[4,5], Jonathan Pierce[3], Haoran Yang[6], M. Zahid Hasan[4,5], Yue Wu[6], Yong P. Chen[1,2,7,*]

[1]Physics department, Purdue University, West Lafayette, IN 47907
[2]Birck Nanotechnology Center, Purdue University, West Lafayette, IN 47907
[3]Center for Solid State Energetics, RTI International, Research Triangle Park, NC 27709
[4]Joseph Henry Laboratories, Department of Physics, Princeton University, Princeton, New Jersey 08544, USA
[5]Princeton Institute for Science and Technology of Materials, Princeton University, Princeton, New Jersey 08544, USA
[6]School of Chemical Engineering, Purdue University, West Lafayette, Indiana 47907
[7]School of Electrical and Computer Engineering, Purdue University, West Lafayette, IN 47907

*Emails: rama@rti.org; yongchen@purdue.edu

**Abstract**: Topological insulator (TI) materials such as $Bi_2Te_3$ and $Bi_2Se_3$ have attracted strong recent interests. Large scale, high quality TI thin films are important for developing TI-based device applications. In this work, structural and electronic properties of $Bi_2Te_3$ thin films deposited by metal organic chemical vapor deposition (MOCVD) on GaAs (001) substrates were characterized via X-ray diffraction (XRD), Raman spectroscopy, angle-resolved photoemission spectroscopy (ARPES), and electronic transport measurements. The characteristic topological surface states (SS) with a single Dirac cone have been clearly revealed in the electronic band structure measured by ARPES, confirming the TI nature of the MOCVD $Bi_2Te_3$ films. Resistivity and Hall effect measurements have demonstrated relatively high bulk carrier mobility of ~350 $cm^2$/Vs at 300K and ~7,400 $cm^2$/Vs at 15 K. We have also measured the Seebeck coefficient of the films. Our demonstration of high quality topological insulator films grown by a simple and scalable method is of interests for both fundamental research and practical applications of thermoelectric and TI materials.

For decades, $Bi_2Te_3$ in its bulk form, and particularly their alloys have been known as one of the best thermoelectric (TE) materials at ordinary temperatures (270 to 400 K) with high figure of merit (ZT ~1)[1,2]. Recently, this material, thought to be a well-studied semiconductor, has been discovered[3,4,5] as a new type of quantum matter, a 3D topological insulator (TI). Due to their many remarkable properties[6,7,8,9,10], topological insulators have currently emerged as one of the most actively researched subjects in condensed matter physics. The bulk of a TI possesses an insulating gap, whereas there exist non-trivial metallic surface states (SS) at the surface resulting from an interplay between the topology of electronic band structure and strong spin-orbit coupling in the bulk. The topological SS are protected by time-reversal symmetry, and therefore cannot be destroyed by perturbations of non-magnetic impurities and small lattice defects. Furthermore, such SS give rise to 2D Dirac fermions with spin-momentum locking and suppressed back scattering, promising a host of novel physics and nano-electronics devices[6,7,8,9,10]. The SS in $Bi_2Te_3$ have been directly revealed by angle-resolved photoemission spectroscopy (ARPES)[4], scanning tunneling microscopy (STM)[11,12], and magneto-transport measurements have shown that the SS carrier mobility could reach as high as ~10,000 $cm^2/Vs$[5].

Synthesizing high quality materials is the foundation for exploiting the unique properties of $Bi_2Te_3$ for TI and thermoelectric device applications. High-quality $Bi_2Te_3$ crystals can be grown by the commonly-used Bridgeman technique[4,5,13], however many applications (e.g. on-chip electronics) desire large scale thin films. Various deposition techniques, such as sputtering[14], evaporation[15,16,17], electrochemical deposition[18], metal organic chemical vapor deposition (MOCVD)[19,20,21] and molecular beam epitaxy (MBE)[22,23], have been developed to grow continuous $Bi_2Te_3$ films on different substrates. Films grown by sputtering, evaporation, and electrochemical deposition, offer room temperature (300 K) carrier mobility that are typically at least 1 order of magnitude lower than that of bulk crystals (~ 300 to 600 $cm^2$/Vs) [13,24,25], and SS have yet to be observed. MBE-grown films have shown mobility ~ 150 $cm^2$/Vs at 300 K[24], and SS have been reported[26]. However, the relatively high cost and low yield of the MBE process can limit its applications in industry. MOCVD has been successfully employed as an industrial method for mass production of thin films and semiconductors. Recent developments in the MOCVD technique have also demonstrated the growth of ultra-short-period superlattices in the $Bi_2Te_3$ system achieving one of the best thermoelectric performance (ZT ~ 2.4 at room temperature)[20]. However, there have been little studies of such MOCVD $Bi_2Te_3$ films from the TI perspective. In this letter, we explore and report on such wafer-scale, high quality $Bi_2Te_3$ thin films grown by MOCVD on GaAs (001) substrates. The TI nature of such films is for the first time directly demonstrated by ARPES. The carrier mobility is ~ 350 $cm^2$/Vs at room temperature and increases to ~ 7,400 $cm^2$/Vs at 15 K, approaching some of the best values reported. We also measured Seebeck coefficients of the film between 220K and 400K. Our results will be important for understanding the structural and electronic properties of $Bi_2Te_3$ films grown by MOCVD and using such films in fundamental research and device applications incorporating thermoelectric power and/or TI effects.

Heteroepitaxial $Bi_2Te_3$ thin films were grown by "van der Waals epitaxy"[21] on GaAs (001) substrates in a vertical, RF-heated, custom-built MOCVD reactor using a novel susceptor design[21]. Prior to loading the GaAs substrates into the reactor chamber, an ex-situ cleaning procedure was performed using solvents (trichloroethylene, acetone, and methanol) for degreasing followed by a piranha etch to remove any remaining contaminants on the growth surface. Trimethylbismuth ($Bi(CH_3)_3$) and diisopropyltelluride ($Te(C_3H_7)_2$) were employed as Bi and Te precursors respectively, transported and diluted using hydrogen gas. A deposition pressure of 350 torr and deposition temperatures between 300°C and 400°C were utilized to achieve heteroepitaxial c-plane orientated $Bi_2Te_3$ films (photograph shown in the inset of Fig. 1b). As-grown films with ~1 μm thickness were cut into smaller pieces for various measurements. We will present the experimental results from the following measurements: high resolution X-ray diffraction (HRXRD) (performed on sample A); Raman spectroscopy (sample B), ARPES (sample C); electronic and thermoelectric transport measurements (resistivity, Hall effect and Seebeck coefficient measurements and their temperature dependence) (sample B). Data taken from similar samples yield qualitatively similar results for all the measurements.

HRXRD was performed using a Phillips X-pert MRD X-ray diffractometer with four-fold Ge (220) monochromator, a three-fold Ge (220) analyzer and Cu Kα radiation (1.5418Å) with the tube energized to 45 keV and 40 mA. The 2-theta-omega coupled scan was measured from 40 to 70 degrees using a 0.02 degree step size. Fig. 1a shows the XRD pattern (measured on sample A) with peaks assigned to the corresponding Miller (*hkl*) indices. The XRD reflections are attributed to $Bi_2Te_3$ (0,0,15) (0,0,18) (0,0,21) planes relative to the cubic GaAs substrate. The absence of additional peaks other than the (00l) family shows that the $Bi_2Te_3$ thin film is indeed grown along the (trigonal) c-axis (i.e., parallel to the $Bi_2Te_3$ quintuple layers). The lattice structure of our sample was further investigated by Raman spectroscopy. Representative Raman spectrum (Fig. 1b) measured (in ambient with a 532 nm excitation laser with circular polarization and ~200 μW incident power) on sample B shows two Raman peaks at 99.8 cm$^{-1}$ and 132.0 cm$^{-1}$, which agree well with the characteristic lattice vibration modes $E_g^2$ and $A_{1g}^2$ observed in previous Raman studies of $Bi_2Te_3$ bulk crystal[27] and thin films[28,29,30].

In order to image the electronic band structure and the topological SS of the MOCVD grown $Bi_2Te_3$, we performed ARPES measurements on the films at the APPLE-PGM beam line of the Synchrotron Radiation Center (SRC), Wisconsin, equipped with a Scienta SES200U electron analyzer. Measurements were performed at ~10 K with 27 eV incident photon energy. Energy resolution was set to be ~ 30 meV. The sample (sample C) was cleaved *in situ* under a vacuum pressure lower than $6 \times 10^{-11}$ torr, and found to be stable and without degradation for the typical measurement period of 24 hours. In the energy-momentum (E-k) band dispersion map (Fig. 2a), a sharp V-shaped (Dirac-like) TI SS band is observed above the valance band. The Dirac point in this image is located at ~240 meV below the Fermi level (the horizontal dashed line in Fig. 2a), which is located inside the bulk band gap as no signature of the bulk conduction band is seen in this image. Detailed analysis of Fig. 2a reveals a Fermi momentum of $k_F$ ~ 0.096 Å$^{-1}$. The Fermi velocity is obtained to be $v_F$ ~ 2.54 eV Å = $3.85 \times 10^5$ m/s, in good agreement with previously measured values for bulk crystals ($3.87 \times 10^5$ m/s)[4] and MBE grown thin films ($3.32 \times 10^5$ m/s)[26]. From the momentum width of the Dirac SS band (FWHM in unit of Å$^{-1}$), we estimate the mean free path $l$ of the SS Dirac fermions to be $l(E_F)$ ~ 47 Å (it should be noted that $l$ is affected by thermal broadening and the limited momentum resolution of the ARPES experimental setup). All parameters are calculated along the $\Gamma - K$ direction. Furthermore, the *surface* carrier concentration is found to be on the order of $7 \times 10^{12}$ cm$^{-2}$ (n-type) as derived from the enclosed area of the SS Fermi surface (Fig. 2b). The Fermi surface of the SS shown in Fig. 2b shows a distinctive deviation from a circular shape, interpreted as due to hexagonal warping, similar to previous observations in $Bi_2Te_3$ bulk crystals[4]. The hexagonal warping that deforms the SS Fermi surface opens up new electronic scattering channels, which could also give rise to exotic physics such as spin-density waves[12,31]. Our ARPES results confirm the topological insulator nature of our high quality MOCVD $Bi_2Te_3$ thin film.

To assess the electronic transport properties of our samples, we characterized sheet resistances and Hall effect using samples of van der Pauw geometry with silver paint contacts. Fig. 3 shows the sheet resistance ($R_s$) of sample B (width × length × thickness ~ 2.5 mm × 5 mm × 1 μm) measured from room temperature down to 15K, displaying a metallic behavior. At room temperature, $R_s$ is 19.7 Ω/□; while at 15 K, it becomes ~16 times smaller with the value of 1.25 Ω/□. The temperature (T) dependence of $R_s$ can be fitted to a simplified phenomenological model[32]

$$R_s = R_0 + \alpha \cdot e^{-\theta/T} + \beta \cdot T^2, \tag{1}$$

where the 3 terms on the right of Eq. (1) correspond to the contributions of impurity scattering (giving rise to a low temperature residual resistance $R_0$), phonon scattering, and electron-electron scattering respectively. We find that $R_0 = 1.39 \pm 0.14$ Ω/□, $\alpha = 12.0 \pm 3.7$ Ω/□, $\theta = 221.8 \pm 33$ K and $\beta = 0.00014 \pm 0.000015$ Ω/□/$K^2$ give the best fit (as shown by the black dash line in Fig. 3) to the experimental data. The temperature dependence is dominated by the exponential term (phonon scattering) and the fitting parameter $\theta$ corresponds to an effective phonon frequency $\omega = \frac{k_B \theta}{\hbar} = 3.2 \times 10^{13}\ rad/s$. The very small value of $\beta$ indicates that e-e scattering effect is weak, as generally expected. All the fitting parameters extracted from our sample are on the similar order of magnitudes with previous values reported for doped bulk $Bi_2Te_3$[32]. The inset of Fig. 3 shows Hall effect of sample B measured at room temperature and 15 K. From the Hall slope, we extract the p-type carrier (hole) density to be $9.1 \times 10^{18}$ $cm^{-3}$ at room temperature and

$6.8 \times 10^{18}$ cm$^{-3}$ at 15K. The comparable high values of hole densities observed between 300K and 15K indicate that our sample is highly p-type doped with a Fermi level located in the valance band. The carrier mobility as extracted from the carrier density and sheet resistance is ~ 350 cm$^2$/Vs at 300 K, and increases to ~ 7,400 cm$^2$/Vs at 15 K. The significant enhancement of the mobility at low temperature is consistent with the freezing out of phonons that give rise to carrier scattering. Both our room temperature and low temperature mobilities approach the highest values previously reported[13,24] in Bi$_2$Te$_3$.

The thermoelectric power (Seebeck coefficient) of the MOCVD grown Bi$_2$Te$_3$ films was measured by a commercial Seebeck measurement system (MMR SB-100). Sample B was sandwiched between a heater and heat sink in vacuum, and the voltage difference was measured between the hot and the cold ends with a maximum temperature fluctuation of ±0.2 K and a voltage resolution of 50 nV. Fig. 4 shows the Seebeck coefficient (S) measured between 220K and 400K. The positive S indicates a p-type carrier, consistent with the positive Hall coefficient (Fig. 3 inset). The Seebeck coefficient peaks at room temperature, reaching a value of ~153 µV/K, comparable with values previously measured in Bi$_2$Te$_3$ at comparable doping levels[25,33].

Due to the highly doped and metallic bulk, the transport properties presented in this study originate dominantly from the bulk carriers and the contribution of SS carriers is negligible. Future strategies to further reduce the bulk conduction and bring out the SS-based TI effects may include various doping engineering[4,34], as well as thinning down the MOCVD films (e.g., to tens of nm). The transport measurements reported in this work, showing a Fermi level located in the valance band does not conflict with our ARPES measurement showing a Fermi level in the band gap. This is related to the fact that the ARPES measured *in-situ* cleaved surfaces under ultra-high vacuum, whereas the transport measurements probe the bulk properties. The bulk band can bend near the surface due to space-charge accumulation, leading to a different Fermi level from its bulk value (our results thus suggest a downward band bending near the sample surface measured in the ARPES) and similar observations have been reported for the (Bi, Sb, Se, Te) family of TI materials[35,36].

In summary, we have synthesized wafer-scale $Bi_2Te_3$ thin films on GaAs (001) substrate by MOCVD, a simple and scalable method. The samples show excellent crystalline characteristics as demonstrated by HRXRD and Raman spectroscopy. The TI nature of our sample, featuring a Dirac-like SS band, has been directly revealed by ARPES. The high carrier mobility further confirms the good electronic quality of the films. Our results demonstrate the potential of MOCVD $Bi_2Te_3$ films in a wide range of applications involving TI and thermoelectrics materials.

We acknowledge support from DARPA MESO program (Grant N66001-11-1-4107). Part of the transport measurements were performed using the user facilities at Argonne National Laboratory (ANL) Center for Nanoscale Materials (CNM), which is supported by U.S. Department of Energy, Office of Science, Office of Basic Energy Sciences, under Contract No. DE-AC02-06CH11357. C. L. acknowledges T. Kondo and A. Kaminski for provision of data analysis software. The Synchrotron Radiation Center is supported by the Natural Science Foundation under Contract No. NSF-DMR-0537588. M. Z. H. acknowledges Visiting Scientist support from LBNL and additional support from the A. P. Sloan Foundation. The authors also thank B. Fisher, Dr. Y. Wang and Prof. X. Xu for experimental assistance and Dr. J. Maassen and Prof. M. Lundstrom for valuable discussions.

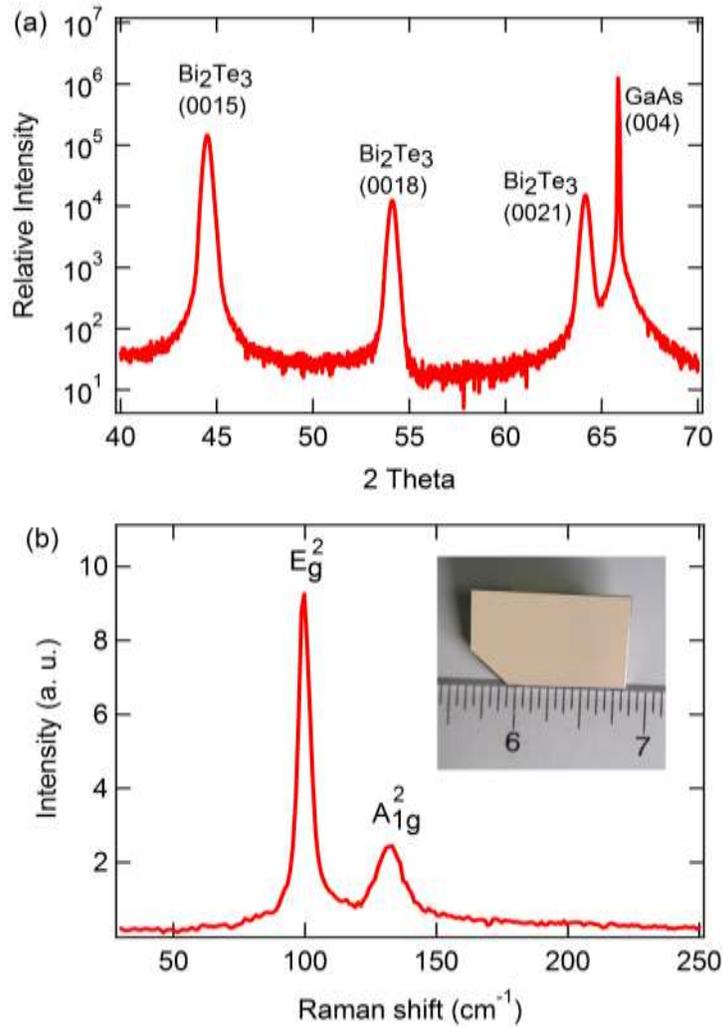

Fig. 1 (a) Representative XRD (X-ray diffraction) pattern on a $Bi_2Te_3$ film grown by MOCVD on GaAs (001) substrate. The peaks are labeled with (*hkl*) indices. (b) Representative Raman spectrum (measured with a 532 nm laser). Two characteristic Raman peaks[27], $E_g^2$ and $A_{1g}^2$, are labeled. Inset shows photograph of a typical as-grown $Bi_2Te_3$ film (thickness ~ 1 μm, as used in this work).

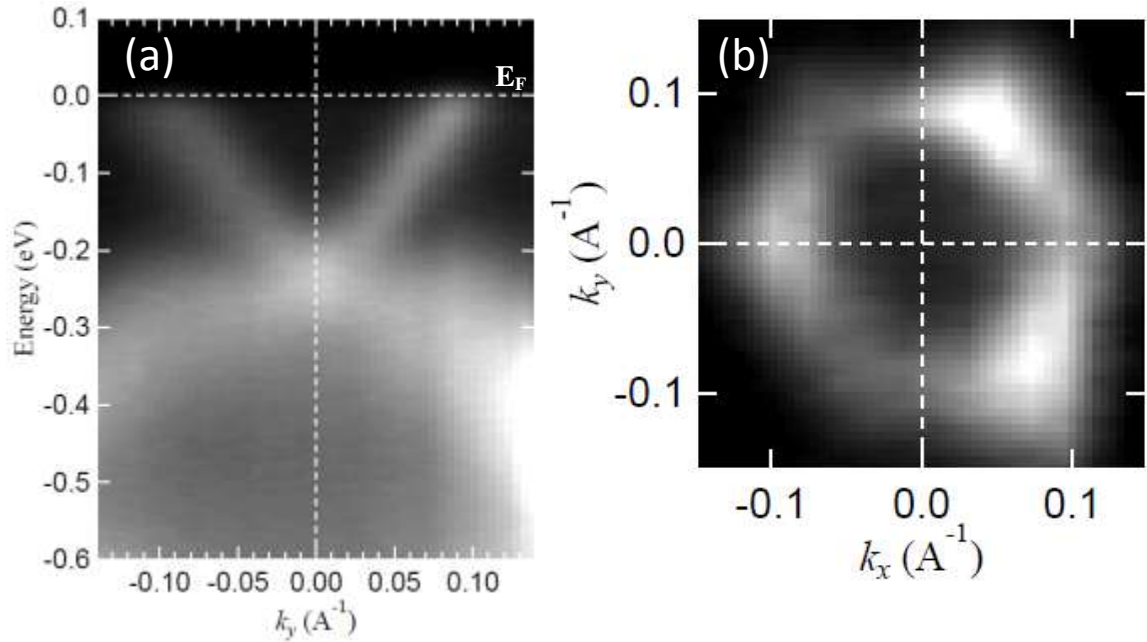

Fig. 2 ARPES (Angle Resolved Photo Emission Spectroscopy) characterization of a $Bi_2Te_3$ thin film grown by MOCVD on GaAs (001). (a) ARPES band dispersion map along the Γ-K direction, clearly revealing the V-shaped topological surface state band. (b) ARPES Fermi surface map. The hexagonal-shaped (due to hexagonal warping[4,31]) of the surface state Fermi surface is clearly revealed.

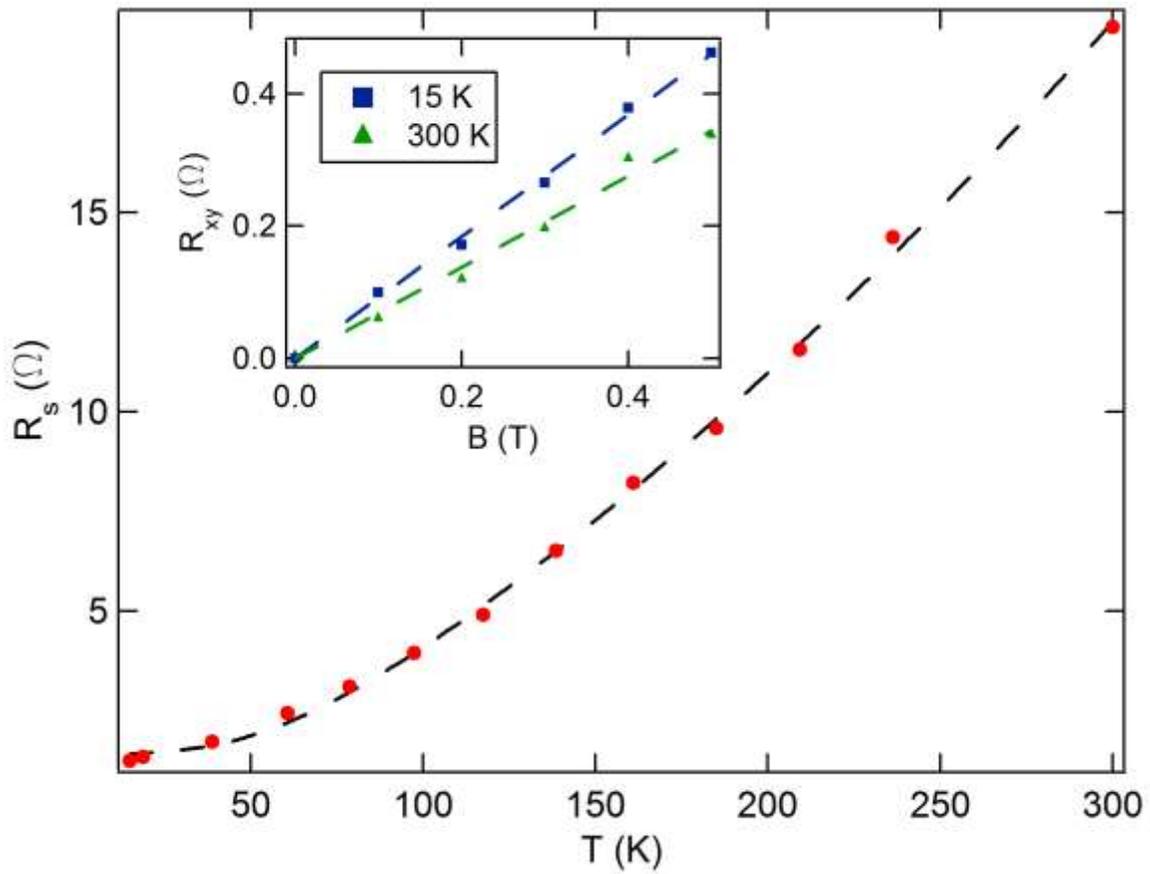

Fig. 3 Temperature dependence of sheet resistance (per square) $R_s$ in sample B (thickness ~ 1 μm), displaying a metallic behavior with $R_s(300K)/R_s(15K)$ ~ 16. The black dash line shows the fitting to Eq. (1). The inset shows Hall resistances $R_{xy}$ of sample B measured at 15 K and 300 K. The dash lines are the linear fittings of the corresponding data.

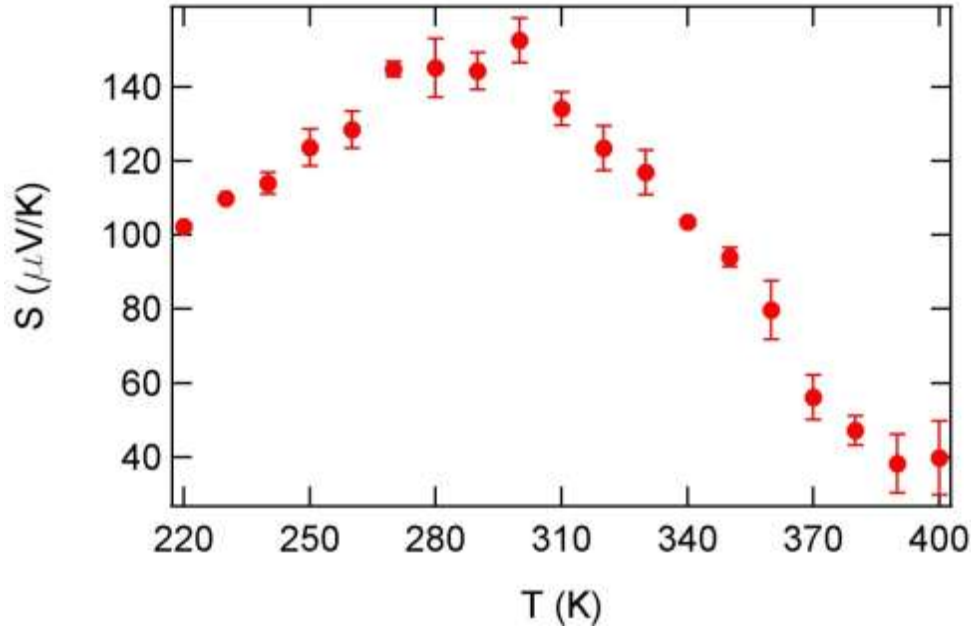

Fig. 4 Seebeck coefficient (S) of sample B measured between 220K and 400K.